\documentstyle[preprint,aps]{revtex}

\begin{document}

\draft

%\preprint{
%\vbox{
%\hbox{UASLP-IF-00-05}
%\hbox{ESFM/F04/00}
%\hbox{hep-ph/00nnmmm}
%}}

\title{
Radiative corrections to the semileptonic Dalitz plot with angular
correlation between polarized decaying and emitted hyperons: effects of
the four-body region
}

\author{
Rub\'en Flores-Mendieta,$^*$ A.~Garc{\'\i}a$^\dagger$,
A.~Mart{\'\i}nez,$^\ddagger$ and J.~J. \ Torres$^\ss$
}

\address{
$^*$Instituto de F{\'\i}sica, Universidad Aut\'onoma de San Luis Potos{\'\i},
\\
\'Alvaro Obreg\'on 64, Zona Centro, San Luis Potos{\'\i}, S.L.P. \ 78000,
Mexico \\
$^{\dagger}$Departamento de F{\'\i}sica, Centro de Investigaci\'on y de
Estudios Avanzados del IPN, \\
Apartado Postal 14-740, M\'exico, Distrito Federal 07000, Mexico \\
$^{\ddagger}$Escuela Superior de F{\'{\i }}sica y Matem\'{a}ticas del IPN, \\
Apartado Postal 75-702, M\'{e}xico, Distrito Federal 07738, Mexico \\
$^{\ss}$Escuela Superior de C\'omputo del IPN, \\
Apartado Postal 75-702, M\'{e}xico, Distrito Federal 07738, Mexico
}

\date{November 6, 2001}

\maketitle

\tightenlines

\begin{abstract}
We obtain a model-independent expression for the complete Dalitz plot of
semileptonic decays of polarized hyperons, which includes both the three-body
and the four-body regions. We calculate radiative corrections to order
$\alpha$, neglecting terms of order $\alpha q/\pi M_{1}$, where $q$ is the
four-momentum transfer and $M_{1}$ is the mass of the decaying hyperon. Our
results exhibit explicitly the correlation between the emitted hyperon
three-momentum and the spin of the decaying hyperon. This allows us to obtain
the corresponding radiative corrections to the integrated emitted hyperon
spin-asymmetry coefficient. Our formulas are valid for charged as well as for
neutral decaying hyperons and are appropriate for model-independent
experimental analysis whether the real photon is discriminated or not.
\end{abstract}


\section{Introduction}

     In previous works~\cite{rfm97,amv00} we have obtained the radiative
corrections (RC) to the Dalitz plot (DP) of hyperon semileptonic decays
(HSD), $A\rightarrow B \ell {\overline \nu}_{\ell}$, for the case of
polarized decaying hyperons ($A$ and $B$ are hyperons and $\ell$ and
$\nu_\ell$ are the accompanying charged lepton and neutrino). In these
calculations we have kept the ${\bf \hat{s}}_1 \cdot {\bf \hat{p}}$ spin
correlation displayed explicitly in the DP. Here ${\bf \hat{s}}_1$ is the spin
of $A$ and ${\bf \hat{p}}$ is a unit vector along the direction of the
three-momentum of either the emitted baryon ${\bf \hat{p}}_2$ or the emitted
charged lepton ${\bf \hat{l}}$. In the former case we can obtain the
spin-asymmetry coefficient of the outgoing baryon~\cite{rfm97}, while in the
latter we can obtain the spin-asymmetry coefficient of the charged
lepton~\cite{amv00}.

     In Ref.~\cite{amv00} we have considered the complete DP including in our
calculations the so-called three- and four-body regions of this DP
(hereafter, these regions will be referred to as TBR and FBR, respectively.)
The results obtained showed us the importance of the contribution of the FBR
to the RC. This region is present when real photons cannot be discriminated in
an experimental analysis of HSD. It is the purpose of this work to extend the
calculations of Ref.~\cite{rfm97} in order to incorporate the four-body
contribution to the RC of the corresponding DP, in the same way as we did in
Ref.~\cite{amv00}. The result will be suitable for model independent analysis
of experiments where real photons cannot be discriminated. We will also
determine the spin-asymmetry coefficient $\alpha_B$ of the outgoing baryon.

     The strategy we follow in order to incorporate the four-body contribution
to the results of Ref.~\cite{rfm97} is the same as the one presented in
Ref.~\cite{amv00}. Accordingly, in Sec.~II we summarize the main results
concerning the RC to the Dalitz plot in the TBR~\cite{rfm97}, and we
rearrange them in parallel to Ref.~\cite{amv00}. In Sec.~III we obtain the
RC to the complete DP in terms of the triple integrals over the photon
bremsstrahlung three-momentum variables, which can be numerically
evaluated. In Sec.~IV we perform analytically these integrals and we give
our second main result, namely, the complete analytical RC to the DP of
decaying polarized hyperons to order $\alpha$ with the ${\bf \hat{s}}_1
\cdot {\bf \hat{p}}_2$ correlation explicitly displayed. In Sec.~V we
obtain the RC to the spin-asymmetry coefficient $\alpha_B$ of the emitted
baryon with the three- and four-body contributions explicitly indicated.
In Sec.~VI we evaluate numerically, for the TBR, the percentage RC to
$\alpha_B$ at very-well defined points of the DP for the decays $\Sigma^-
\rightarrow n e{\overline \nu}$ and $\Lambda \rightarrow pe{\overline
\nu}$. We also evaluate for these two decays, the percentage ratio of the
spin-dependent part of the DP to their spin independent part at different
points of the FBR. The RC to the integrated spin-asymmetry coefficient
$\alpha_B$ are also evaluated for several decays. All these results are
compared with those of Ref.~\cite{gluck} and we find that the agreement is
acceptable. Finally, in Sec.~VII we present our conclusions.

\section{TBR radiative corrections to the DP}

     In this section we shall first briefly review the results of
Ref.~\cite{rfm97}, without repeating details that can be found there, and we
shall introduce our notation. Second, it turns out that in order to study the
FBR, it is convenient to follow the steps of Ref.~\cite{amv00}. This then
requires to rearrange the expressions of Ref.~\cite{rfm97}, before the photon
three-momentum is integrated, and cast them into a new form which keeps a
close parallelism with Ref.~\cite{amv00}. Doing this will make the analysis of
the FBR very expedient and transparent, as will be appreciated in the next
section.

     For definiteness, let us consider the HSD
\begin{equation}
A \rightarrow B + \ell + {\overline \nu}_{\ell}, \label{eq:dec}
\end{equation}
where the lepton $\ell$ is negatively charged. How to extend our results
to the case when $\ell$ is positively charged will be discussed in Sec.
IV. Our notation is the same as before \cite{rfm97,amv00}. Thus $p_1 =
\left(E_1, {\bf p}_1 \right)$, $p_2 = \left(E_2, {\bf p}_2 \right)$, $l =
\left(E, {\bf l} \right)$, and $p_\nu^0 = \left(E_\nu^0, {\bf p}_\nu
\right)$ are the four-momenta of $A$, $B$, $\ell$, and ${\overline
\nu}_\ell$, respectively. $M_1$, $M_2$, and $m$ are the non-zero masses of
the first three particles. In the center-of-mass frame of $A$, the
quantities $p_2$, $l$, and $p_\nu$ will also denote the magnitudes of the
corresponding three-momenta. All other conventions and notation are given
in Ref.~\cite{rfm97}.

     The result for the virtual RC to process~(\ref{eq:dec}) is compactly
given by Eq.~(15) of Ref.~\cite{rfm97}, namely,

\begin{eqnarray}
 d\Gamma_V & = & d \Omega \left\{A_0^\prime + \frac{\alpha}{\pi} \,
(A_1^\prime \, \phi + A_1^{\prime \prime} \, \phi^\prime) - {\bf \hat s}_1
\cdot {\bf \hat p}_2 \left[A_0^{\prime \prime} + \frac{\alpha}{\pi} \, ( \,
A_2^\prime \, \phi + A_2^{\prime \prime} \, \phi^\prime \, ) \right] \right\},
\label{eq:virtual}
\end{eqnarray}
where
\begin{equation}
d \Omega = \frac{G_V^2}{2} \, \frac{dE_2 \, dE \,
d\Omega_2}{{(2\pi)}^4} \, 2M_1 . \label{eq:domega}
\end{equation}
There is no need to reproduce here the detailed expressions of the
contributions in Eq.~(\ref{eq:virtual}), so we only provide the guide to
find them. Respectively, $A_0^\prime$, $A_1^\prime$, $\phi$, $A_1^{\prime
\prime}$, $\phi^\prime$, $A_0^{\prime \prime}$, $A_2^{\prime}$, and
$A_2^{\prime \prime}$ are given by Eqs.~(16), (17), (8), (18), (9), (19),
(20), and (21) of Ref.~\cite{rfm97}.

     As for the bremsstrahlung contribution, the approach to compute RC
to the DP is discussed in full in Refs.~\cite{rfm97,amv00}, so only a
few salient facts will be repeated here. We need to consider the
four-body decay
\begin{equation}
A \rightarrow B + \ell + {\overline \nu}_{\ell} + \gamma, \label{eq:bre}
\end{equation}
where $\gamma$ represents a real photon with four-momentum $k = (w,{\bf
k})$. The TBR of the DP is the region where the three-body
decay~(\ref{eq:dec}) and the four-body decay~(\ref{eq:bre}) overlap
completely. The FBR is where in process~(\ref{eq:bre}) neither of
the energies of ${\overline \nu}_{\ell}$ and $\gamma$ can be made
zero. The complete DP can be seen as the union of the TBR and the FBR. The
bounds for the kinematical variables in both regions are defined in
Ref.~\cite{amv00}.

     The differential decay rate for process~(\ref{eq:bre}), given by Eq.~(32)
of Ref.~\cite{rfm97}, reads
\begin{equation}
d\,\Gamma_B^{{\rm TBR}} = d\,\Gamma_B^{\prime {\rm TBR}} - d \,
\Gamma_B^{(s)\,{\rm TBR}} , \label{eq:dectbr}
\end{equation}
where $d\,\Gamma_B^{\prime \,{\rm TBR}}$ and $d \, \Gamma_B^{(s)\,{\rm
TBR}}$ are the spin-independent and spin-dependent contributions of
$d\,\Gamma_B^{{\rm TBR}}$, respectively. Unlike Eq.~(32) of
Ref.~\cite{rfm97}, here we have added the superscript TRB to the several
quantities in Eq.~(\ref{eq:dectbr}) to emphasize the fact that they are
defined in the TBR only, a distinction that is now necessary.
$d\,\Gamma_B^{\prime \,{\rm TBR}}$ is given in Eq.~(33) of
Ref.~\cite{rfm97} with the explicit forms for its contributions in
Eqs.~(34), (37), and (38). Similarly, $d \, \Gamma_B^{(s)\,{\rm TBR}}$ is
given by Eq.~(43) of this reference with Eqs.~(51) and (57) for its
corresponding explicit contributions. For our present purposes it is
convenient to rearrange these equations into the forms introduced in
Ref.~\cite{amv00}. Therefore, $d\,\Gamma_B^{\prime \,{\rm TBR}}$ becomes
\begin{equation}
d\,\Gamma_B^{\prime \,{\rm TBR}} = \frac{\alpha}{\pi} d \Omega
\left\{A_1^\prime I_0(E,E_2) + \frac{p_2 l}{4\pi} \int_{-1}^1 dx
\int_{-1}^{y_0} dy \int_0^{2\pi} d \varphi_k \left[\left| {\sf M}^\prime
\right|^2 + \left| {\sf M}^{\prime \prime} \right|^2 \right] \right\},
\label{eq:spinindtbr}
\end{equation}
which agrees with Eq.~(27) of this Ref.~\cite{amv00}. Whereas $d \,
\Gamma_B^{(s)\,{\rm TBR}}$ becomes
\begin{equation}
d\,\Gamma_B^{(s) \,{\rm TBR}} = \frac{\alpha}{\pi} d\Omega
{\bf \hat{s}}_1 \cdot {\bf \hat{p}}_2 \left\{ A_2^\prime I_0(E,E_2) +
\frac{p_2l}{4\pi} \int_{-1}^1 dx \int_{-1}^{y_0} dy
\int_0^{2\pi} d\varphi_k \left[ \left| {\sf N}^{\prime \prime \prime}
\right|^2 + \left| {\sf N}^{{\rm IV}} \right|^2 \right] \right\}.
\label{eq:spindeptbr}
\end{equation}
Here $I_0(E,E_2)$, given by Eq.~(52) of Ref.~\cite{rfm97}, fully
contains the infrared divergence which will be canceled by its
counterpart contained in the virtual RC to the differential decay rate,
namely, Eq.~(\ref{eq:virtual}). The quantities $\left| {\sf M}^\prime
\right|^2$ and $\left| {\sf M}^{\prime \prime} \right|^2$ are explicitly given
by Eqs.~(28) and (29) of Ref.~\cite{amv00}. Whereas $\left| {\sf
N}^{\prime \prime \prime} \right|^2$ and $\left| {\sf N}^{{\rm IV}}
\right|^2$ are new, namely,
\begin{eqnarray}
\left| {\sf N}^{\prime \prime \prime} \right|^2 = \frac{\beta^2}{p_2}
\left[D_3 \left(E_\nu^0 + \frac{p_2ly}{D} \right) - D_4 E \left(1 -
\frac{{\bf p}_2 \cdot {\bf {\hat k}}}{D} \right) \right]
\frac{1-x^2}{{(1-\beta x)}^2} \label{eq:n3}
\end{eqnarray}
and
\begin{eqnarray}
\left| {\sf N}^{{\rm IV}} \right|^2 & = & \frac{1}{ED(1-\beta x)} \left[
- D_3 E_\nu \left(w + E - \frac{m^2}{E(1-\beta x)} \right) {\bf {\hat
p}}_2 \cdot {\bf {\hat k}} - D_3 E_\nu ly \right. \nonumber \\
&   & \mbox{\hglue1.4truecm} + \left. D_4 \left(w + E(1+\beta x) -
\frac{m^2}{E(1-\beta x)} \right) (ly+p_2+w {\bf {\hat p}}_2 \cdot {\bf
{\hat k}}) \right] . \label{eq:n4}
\end{eqnarray}
The counterparts of $\left| {\sf N}^{\prime \prime \prime} \right|^2$ and
$\left| {\sf N}^{{\rm IV}} \right|^2$ in Ref.~\cite{amv00} are $\left|
{\sf M}^{\prime \prime \prime} \right|^2$ and $\left| {\sf M}^{{\rm IV}}
\right|^2$, respectively, given there by Eqs.~(42) and (43).

     In these last equations, $y = {\bf \hat l}\cdot{\bf \hat p}_2$, $x
= {\bf \hat l}\cdot{\bf \hat k}$, $D = E_\nu^0 + ({\bf l} + {\bf p}_2)
\cdot {\bf \hat k}$, and $E_\nu = E_\nu^0 - w$, where $E_\nu^0$ is the
neutrino energy available when the photon is not present in the decay,
$\beta = l/E$, and $\varphi_k$ is the azimuthal angle of the real photon.
The $D_i$ are quadratic functions of the leading form factors, they are
introduced in those Eqs.~(42) and (43), and are explicitly given in Eqs.~(B13)
and (B14) of the same reference.

     Adding Eqs.~(\ref{eq:virtual}) and (\ref{eq:dectbr}) we obtain the
differential decay rate with RC for the TBR, corresponding to Eq.~(100) of
Ref.~\cite{rfm97}, but it is now rearranged in parallelism with Eq.~(44) of
Ref.~\cite{amv00}. This decay rate has the real photon three-momentum
integrations ready to be performed numerically [see Eqs.~(\ref{eq:spinindtbr})
and (\ref{eq:spindeptbr})].

     To conclude our short review of Ref.~\cite{rfm97}, we must mention
that the photon three-momentum integrations can be performed analytically.
The result is the one given in Eq.~(96) of Ref.~\cite{rfm97}, namely,
\begin{eqnarray}
d\Gamma_B^{\rm TBR} & = & \frac{\alpha}{\pi} d \Omega \{(D_1 + D_2)
(\theta^\prime + \theta^{\prime \prime \prime}) + D_2 (\theta^{\prime
\prime} + \theta^{\rm IV}) + A_1^\prime \theta_1 \nonumber \\
& & \mbox{\hglue2.0truecm} - {\bf \hat s}_1 \cdot {\bf \hat p}_2 \, [ \,
A_2^\prime \theta_1 + D_3 ( \rho_1 + \rho_3 ) + D_4 (\rho_2 + \rho_4) \,
] \, \} . \label{eq:rbre}
\end{eqnarray}
There is no need to repeat here the detailed expressions of the quantities
that appear in Eq.~(\ref{eq:rbre}). $\theta_1 = I_0(E,E_2)$, the $\rho_i$
are given in Eqs.~(75)--(78), and $\theta^\prime + \theta^{\prime \prime
\prime}$ and $\theta^{\prime \prime} + \theta^{\rm IV}$ are given in Eqs.
(97) and (98) of this reference. However, an erratum was detected in
$\theta^\prime + \theta^{\prime \prime \prime}$ and it was corrected in
Ref.~\cite{amv00}. One should better use Eqs.~(B39) and (B40) of this
last reference for $\theta^\prime + \theta^{\prime \prime \prime}$ and
$\theta^{\prime \prime} + \theta^{\rm IV}$.

     Collecting partial results, Eqs.~(\ref{eq:virtual}) and
(\ref{eq:rbre}), we obtain for Eq.~(100) of Ref.~\cite{rfm97} the
analytical DP of HSD with non-zero polarization of the initial hyperon
including RC to order $\alpha$ and restricted to the TBR. It is given by
Eq.~(101) of this reference and reads
\begin{eqnarray}
d \Gamma^{\rm TBR} (A \rightarrow B l {\overline \nu}_\ell) & = & d
\Omega \left\{A_0^\prime + \frac{\alpha}{\pi} \, \Phi_1 - {\bf \hat s}_1
\cdot {\bf \hat p}_2 \left[A_0^{\prime \prime} + \frac{\alpha}{\pi} \,
\Phi_2\right] \right\} , \label{eq:dwp}
\end{eqnarray}
where $\Phi_1$ and $\Phi_2$ can be found in Eqs.~(102) and (103) of this
same reference.

\section{FBR bremsstrahlung and complete RC}

     We now come to the main issue of this paper, to obtain the contributions
of the FBR to the RC of the decay (\ref{eq:dec}). It is here where the effort
of the last section, putting the bremsstrahlung contributions of
Ref.~\cite{rfm97} in close parallelism with their counterparts in
Ref.~\cite{amv00}, comes to our advantage. The calculation can now be
performed following the same steps of Sec.~III-C of this last reference. It is
not necessary to repeat here the details. The point is that there it is shown
that the FBR bremsstrahlung differential decay rate has the same structure as
the TBR one, Eq.~(\ref{eq:dectbr}). Namely,
\begin{equation}
d\,\Gamma_B^{{\rm FBR}} = d\,\Gamma_B^{\prime \,{\rm FBR}} - d \,
\Gamma_B^{(s)\,{\rm FBR}} , \label{eq:decfbr}
\end{equation}
where $d\,\Gamma_B^{\prime \,{\rm FBR}}$ and $d \, \Gamma_B^{(s)\,{\rm
FBR}}$ are again the spin-independent and spin-dependent contributions.

    Now, instead of Eqs.~(\ref{eq:spinindtbr}) and (\ref{eq:spindeptbr})
we get explicitly
\begin{equation}
d\,\Gamma_B^{\prime \, {\rm FBR}} = \frac{\alpha}{\pi} d \Omega
\left\{A_1^\prime I_{0F}(E,E_2) + \frac{p_2 l}{4\pi} \int_{-1}^1 dx
\int_{-1}^1 dy \int_0^{2\pi} d \varphi_k \left[\left| {\sf M}^\prime
\right|^2 + \left| {\sf M}^{\prime \prime} \right|^2 \right] \right\}
\label{eq:spinindfbr}
\end{equation}
and
\begin{equation}
d\,\Gamma_B^{(s) \,{\rm FBR}} = \frac{\alpha}{\pi} d\Omega
{\bf \hat{s}}_1 \cdot {\bf \hat{p}}_2 \left\{ A_2^\prime I_{0F} \left(
E,E_2 \right) + \frac{p_2l}{4\pi} \int_{-1}^1 dx \int_{-1}^1 dy
\int_0^{2\pi} d\varphi_k \left[ \left| {\sf N}^{\prime \prime \prime}
\right|^2 + \left| {\sf N}^{{\rm IV}} \right|^2 \right] \right\}.
\label{eq:spindepfbr}
\end{equation}

     The changes between Eqs.~(\ref{eq:spinindtbr})--(\ref{eq:spindeptbr})
and Eqs.~(\ref{eq:spinindfbr})--(\ref{eq:spindepfbr}) are very simple. The
upper limit $y_0$ of Eqs.~(\ref{eq:spinindtbr})--(\ref{eq:spindeptbr})
becomes one in Eqs.~(\ref{eq:spinindfbr})--(\ref{eq:spindepfbr}) and the
infrared divergent $I_0=(E,E_2)$ becomes the infrared convergent
$I_{0F}(E,E_2)$, which is explicitly given in Eq.~(37) of Ref.~\cite{amv00}.
Everything else in these equations is the same. The result
Eq.~(\ref{eq:decfbr}) exhibits only the angular correlation ${\bf \hat
s}_1 \cdot {\bf \hat p}_2$. The other two angular correlations ${\bf
\hat{s}}_1 \cdot {\bf \hat k}$ and ${\bf \hat{s}}_1 \cdot {\bf \hat l}$
were eliminated in favor of the former one using the replacement~\cite{gluck}
\begin{equation}
{\bf \hat s}_1 \cdot {\bf p} \to ({\bf \hat s}_1 \cdot {\bf \hat p}_2)
({\bf p} \cdot {\bf \hat p}_2),
\end{equation}
with ${\bf p} = {\bf l}, \, {\bf k}$. In Ref.~\cite{amv00}, it was the
angular correlation ${\bf \hat{s}}_1 \cdot {\bf \hat l}$ that was extracted.
The counterpart of the present Eq.~(\ref{eq:decfbr}) is Eq.~(39) of that
reference. Both equations have the same form as mentioned above and the
detailed changes are that $d \Omega^\prime$, $\left| {\sf M}^{\prime
\prime \prime} \right|^2$, and $\left| {\sf M}^{\rm IV} \right|^2$ of that
Eq.~(39) are now replaced by Eqs.~(\ref{eq:domega}), (\ref{eq:n3}),
and (\ref{eq:n4}) in Eq.~(\ref{eq:decfbr}). Of course, the
spin-independent $d\Gamma_B^{\prime\, \rm FBR}$ of that Eq.~(39) and this
Eq.~(\ref{eq:decfbr}) is the same, except for the change of $d
\Omega^\prime$ into $d \Omega$.

     The complete RC to process (\ref{eq:dec}) without the restriction
of eliminating real photons (either by direct detection or indirect
energy-momentum conservation) is given by the addition of
Eqs.~(\ref{eq:virtual}), (\ref{eq:dectbr}), and (\ref{eq:decfbr}). The
result can be compactly written as
\begin{equation}
d \Gamma \left( A \rightarrow B \ell {\overline \nu}_\ell \right) =
d\Gamma^{\rm TBR} + d \Gamma^{\rm FBR}. \label{eq:decayrate}
\end{equation}

     This equation is our first main result. The correlation ${\bf \hat
s}_1 \cdot {\bf \hat p}_2$ is explicitly extracted and the integral over the
photon variables ($\varphi_k$, $y$, and $x$) are ready to be performed
numerically. However, all the photon integrals can be analytically performed.
Those of the TBR were already computed before and we reviewed the result in
Eq.~(\ref{eq:dwp}) of Sec.~II. In the next section we shall obtain the
analytical result for the new photon integrals that appear in the FBR
contributions. This will lead to our second main result.

\section{Analytical Integrations}

     Let us now proceed to obtain the analytical expression of
Eq.~(\ref{eq:decfbr}). Not all of the photon integrals of the FBR in
Eqs.~(\ref{eq:spinindfbr}) and (\ref{eq:spindepfbr}) are new. It turns out
that those of $d\Gamma_B^{\prime\, \rm FBR}$ were already performed in
Ref.~\cite{amv00}, which as explained in the last section was to be
expected. All we have to do in this respect is to bring here the result of
Ref.~\cite{amv00}. That is,
\begin{equation}
d \Gamma_B^{\prime \, \rm FBR} = \frac{\alpha}{\pi} d\Omega \left[
A_1^\prime I_{0F}(E,E_2) + (D_1+D_2) \left( \theta_F^\prime +
\theta_F^{\prime
\prime \prime} \right) + D_2 \left(\theta_F^{\prime \prime} +
\theta_F^{\rm IV} \right) \right] .
\end{equation}

     On the other hand, the photon integrals in Eq.~(\ref{eq:spindepfbr})
are new. Their calculation is quite straightforward, albeit tedious. It
is, however, important to invest some extra effort to rearrange their
result so that the notation resembles and uses as-much-as possible
expressions already defined. Without giving further details, the result is
\begin{equation}
d\Gamma_B^{(s)\, \rm FBR} = \frac{\alpha}{\pi} d\Omega {\bf \hat s}_1
\cdot {\bf \hat p}_2 \left[ A_2^\prime I_{0F}(E,E_2) + D_3 \left(
\rho_{1F} + \rho_{3F} \right) + D_4 \left(\rho_{2F} + \rho_{4F} \right)
\right] .
\end{equation}

    The functions $\rho_{iF}$ have the same structure as the $\rho_i$
previously defined in Ref.~\cite{rfm97} for the TBR. Their explicit
expressions are
\begin{eqnarray}
\rho_{1F} & = & \frac{l}{2} [ \, 2 {E_\nu}^0 \theta_{0F} - \zeta_{10F} +
2 \zeta_{11F} + (\beta^2 - 1) \zeta_{12F} \, ] \, , \label{eq:rhoone} \\
\rho_{2F} & = & \frac{E}{2} [ \, - 2l \theta_{0F} - \chi_{10F} + 2
\chi_{11F} + (\beta^2 - 1) \chi_{12F} \, ] \, , \label{eq:rhotwo} \\
\rho_{3F} & = & \frac{\beta}{4} [ - 2 {E_\nu}^0 \zeta_{11F} +
\zeta_{21F}] \nonumber + \frac{1}{2} \left[ - {E_\nu}^0 \chi_{11F} +
\frac{E - {E_\nu}^0}{2E} \chi_{21F} + \frac{\chi_{31F}}{4 E}
\right. \nonumber \\
&   & \mbox{} + \left. (1 - \beta^2) {E_\nu}^0 \chi_{12F} - \frac{1}{2}
( 1 - \beta^2) \chi_{22F} \right] , \label{eq:rhothree} \\
\rho_{4F} & = & \frac{\beta}{2} \left[ -E \zeta_{10F} + 2 E \zeta_{11F}
+ \frac{\zeta_{21F}}{2} - \frac{m^2}{E} \zeta_{12F} \right] +
\frac{p_2^2 \beta}{2} \gamma_{0F} \nonumber \\
&   & \mbox{} + \frac{1}{4} \left[ - \chi_{20F} + 2 \chi_{21F} +
\frac{\chi_{31F}}{2 E} - \frac{m^2}{E^2} \chi_{22F} \right] .
\label{eq:rhofour}
\end{eqnarray}
The structure of the functions $\chi_{iF}$ and $\zeta_{iF}$ is exactly the
same as their counterparts for the TBR. To obtain them explicitly all one
needs to do is replace the $\theta_i$ and $\eta_i$ that appear in the latter
by $\theta_{iF}$ and $\eta_{iF}$. The $\theta_{iF}$ were already given in
Appendix B of Ref.~\cite{amv00}. The $\eta_{iF}$ are new. Before performing
the last analytical integration they are given by
\begin{equation}
\eta_{0F} = \int_{-1}^1 dy,
\end{equation}
\begin{equation}
\eta_{1F} = \int_{-1}^1 dy \frac{1}{G(y)} ,
\end{equation}
\begin{equation}
\eta_{(2+j)F} = \int_{-1}^1 dy [G(y)]^{1/2 - j} \ln {\left[ \frac{E_\nu^0
+ [G(y)]^{1/2}}{E_\nu^0 -  [G(y)]^{1/2}} \right]},
\end{equation}
where $j=0,1,2$ and
\begin{equation}
G(y) = {E_\nu^0}^2 + 2p_2l(y-y_0) .
\end{equation}

     After performing the $y$ integration, their explicit forms are
\begin{equation}
\eta_{1F} = \frac{1}{2p_2l} \ln{\left[ \frac{(p_2+l)^2}{(p_2-l)^2}
\right]} ,
\end{equation}
\begin{eqnarray}
\eta_{2F} & = & \frac{1}{3p_2l} \left\{4E_\nu^0p_2l + {E_\nu^0}^3 \ln{
\left[\frac{y_0-1}{y_0+1} \right]} + (p_2+l)^3 \ln{ \left[
\frac{E_\nu^0+p_2+l}{E_\nu^0-p_2-l} \right]} \right. \nonumber \\
&  & \mbox{\hglue1.0truecm} - \left. (p_2-l)^3
\ln{\left[\frac{E_\nu^0+p_2-l}{E_\nu^0-p_2+l} \right]} \right\} ,
\end{eqnarray}
\begin{eqnarray}
\eta_{3F} & = & \frac{1}{p_2l} \left\{ E_\nu^0 \ln{ \left[
\frac{y_0-1}{y_0+1} \right]} + (p_2+l) \ln{\left[ \frac{E_\nu^0 +
p_2 + l}{E_\nu^0 - p_2 - l} \right]} \right.
\nonumber \\
&  & \mbox{\hglue1.0truecm} - \left. (p_2-l) \ln{\left[ \frac{E_\nu^0 +
p_2 - l}{E_\nu^0 - p_2 + l} \right]} \right\} ,
\end{eqnarray}
\begin{eqnarray}
\eta_{4F} & = & \frac{1}{E_\nu^0 p_2l} \left\{ \ln{ \left[
\frac{(p_2+l)^2}{(p_2-l)^2} \right]} - \ln{ \left[ \frac{y_0-1}{y_0+1}
\right]} + \frac{E_\nu^0}{p_2-l}
\ln{ \left[ \frac{E_\nu^0 + p_2 -l}{E_\nu^0 -p_2+l} \right]}
\right. \nonumber \\
&  & \mbox{\hglue1.0truecm} - \left. \frac{E_\nu^0}{p_2+l} \ln{ \left[
\frac{E_\nu^0 + p_2 +l}{E_\nu^0 -p_2-l} \right]} \right\} ,
\end{eqnarray}
At this point it is convenient to mention that all the $\rho _{iF}\,$
functions are convergent when $p_2 \rightarrow 0$. Therefore, the value of
the bremsstrahlung RC when $E_2 \rightarrow M_2$ is finite and their
numerical evaluations present no problems.

     Following Eq.~(\ref{eq:decfbr}), the FBR analytical bremsstrahlung
differential decay rate of decaying polarized
hyperons reads
\begin{equation}
d\Gamma_B^{\rm FBR} = \frac{\alpha}{\pi} d \Omega \left[\Phi_{1F} - {\bf
\hat{s}}_1 \cdot {\bf \hat{p}}_2 \; \Phi_{2F} \right] , \label{eq:dfbr}
\end{equation}
with
\begin{eqnarray}
\Phi _{1F} & = & A_1^\prime I_{0F}(E,E_2) + \left(D_1 + D_2 \right)
\left( \theta_F^\prime + \theta_F^{\prime \prime \prime} \right) + D_2
\left( \theta_F^{\prime \prime} + \theta_F^{{\rm IV}} \right) ,
\label{eq:e66} \\
\Phi_{2F} & = & A_2^\prime I_{0F}(E,E_2) + D_3 \left(\rho_{1F} +
\rho_{3F} \right) + D_4 \left(\rho_{2F} + \rho_{4F} \right) .
\label{eq:e67}
\end{eqnarray}

     The complete analytical RC to the DP of polarized decaying hyperons
to order $\alpha$, in the approximation of neglecting terms of order
$\alpha q/\pi M_1$, is obtained by adding $d \Gamma^{\rm TBR}$ and $d
\Gamma_B^{\rm FBR}$, Eqs.~(\ref{eq:dwp}) and (\ref{eq:dfbr}). The final
expression is
\begin{equation}
d\Gamma_{\rm TOT} = \frac{G_V^2}{2} \frac{dE_2 \, dE \,
d\Omega_2}{{(2\pi)}^4} 2M_1 \left\{A_0^\prime + \frac{\alpha}{\pi}
\left(\Phi_1 + \Phi_{1F} \right) - {\bf \hat{s}}_1 \cdot {\bf
\hat{p}}_2 \left[ A_0^{\prime \prime} + \frac{\alpha}{\pi} \left(\Phi_2
+ \Phi_{2F} \right) \right] \right\} . \label{eq:dtotal}
\end{equation}

     This is the analytical counterpart of the DP of
Eq.~(\ref{eq:decayrate}) and our second main result. In the next section
we will use this Eq.~(\ref{eq:dtotal}) in order to obtain the
spin-asymmetry coefficient of the emitted baryon, $\alpha_B$.

    Equations (\ref{eq:decayrate}) and ~(\ref{eq:dtotal}) were obtained for the
case when the emitted lepton $\ell$ is negatively charged. The expressions for
the case when $\ell$ is positively charged are obtained \cite{torres} by
changing the sign of each axial form factor $g_i$ $(i=1,2,3)$ and by reversing
the sign in front of ${\bf \hat s}_1 \cdot {\bf \hat p}_2$, in these equations.

\section{Spin-asymmetry coefficient $\alpha_B$}

     Here we will discuss the total RC of order $\alpha$ to the
spin-asymmetry coefficient $\alpha_B$ of the outgoing baryon. For this
purpose, we will use the complete DP with RC, Eq.~(\ref{eq:dtotal}), in
order to calculate the quantities $N^\pm$ in terms of which $\alpha_B$ is
defined, namely,
\begin{equation}
\alpha_B = 2 \frac{N^+ - N^-}{N^+ +N ^-} . \label{eq:npnm}
\end{equation}

     $N^+$ $(N^-)$ denotes the number of baryons with momenta ${\bf
\hat p}_2$ emitted in the forward (backward) hemisphere with respect to
the polarization of the decaying hyperon. Thus, $\alpha_B$ can be
written as
\begin{equation}
\alpha_B^T = -\frac{B_2 + (\alpha/\pi) \left(a_2 + a_{2F} \right)}{B_1 +
(\alpha /\pi) \left( a_1 + a_{1F} \right) }. \label{eq:alphaT}
\end{equation}
In this equation the superscript $T$ (for total) on $\alpha_B$ indicates
that the contributions of both the TBR and the FBR are taken into account.
$B_2$ and $B_1$ are given by Eqs.~(109) and (108) of Ref.~\cite{rfm97}.
The RC to the spin-asymmetry parameter corresponding only to the TBR are
obtained by setting $a_{1F} = a_{2F}=0$ in Eq.~(\ref{eq:alphaT}). In
this case the $\alpha_B$ parameter is
\begin{equation}
\alpha_B^R = - \frac{B_2 + (\alpha/\pi) a_2}{B_1 + (\alpha/\pi) a_1},
\label{eq:alphaR}
\end{equation}
where the superscript $R$ (for restricted) denotes the TBR contribution
only. The uncorrected angular spin-asymmetry coefficient of the emitted
hyperon is simply given by
\begin{equation}
\alpha_B^0 = -\frac{B_2}{B_1} . \label{eq:alpha0}
\end{equation}
$a_1$ and $a_2$ are defined in Eqs.~(112) and (113) of Ref.~\cite{rfm97}.
$a_{1F}$ and $a_{2F}$ are the new FBR contributions to the RC. Explicitly,
they are defined using Eq.~(\ref{eq:dfbr}) as
\begin{equation}
a_{iF} = \int_m^{E_B} \int_{M_2}^{E_2^-} \Phi_{iF} dE_2 dE,
\end{equation}
where $i=1,2$. The kinematical bounds of the FBR are $E_B$ and $E_2^-$ and
they are given in Eqs.~(20) and (16) of Ref.~\cite{amv00}, respectively.

     Equation~(\ref{eq:alphaT}) can be further expanded and can be rewritten in
such a way that only terms of order $\alpha$ appear. Neglecting terms of
order $\alpha q/\pi M_1$ we rearrange $\alpha_B^T$ as
\begin{equation}
\alpha_B^T = \alpha_B^0 \left[1 + \frac{\alpha}{\pi} \left(\frac{a_2
+a_{2F}}{B_2(0)} - \frac{a_1 + a_{1F}}{B_1(0)} \right) \right],
\label{eq:aapprox}
\end{equation}
with
\begin{equation}
B_i(0) = \int_m^{E_m} \int_{E_2^-}^{E_2^+} A_i^\prime dE_2 dE.
\end{equation}
$E_m$ is the maximum energy of the electron and $E_2^+$ is
the upper boundary of the TBR in the DP. Their explicit forms appear in
Eqs.~(17) and (16), respectively, of Ref.~\cite{amv00}.

    A word of caution is necessary here. Equation~(\ref{eq:aapprox}) may be
employed provided $\left| B_2 - B_2 (0) \right| \ll \left| B_2 (0) \right|$.
It may happen that this condition is not met when certain values of the
leading form factors are assumed. This anomalous situation occurs when $f_1(0)
\approx 0$, as is the case in $\Sigma^\pm \to \Lambda e \overline{\nu}$. In
fact, one can show either analytically or numerically that $B_2(0) \approx 0$
when $f_1(0) \approx 0$ and, accordingly, Eq.~(\ref{eq:aapprox}) becomes
ill-defined. When this occurs we should only use the unexpanded version
Eq.~(\ref{eq:alphaT}). When $f_1(0)$ is appreciably large, the results
obtained with Eqs.~(\ref{eq:aapprox}) and (\ref{eq:alphaT}) are acceptable
within our approximations. In case of doubt it is safer to simply use
Eq.~(\ref{eq:alphaT}).

     In Ref.~\cite{rfm97} we only obtained $\alpha_B$ corresponding to the
TBR. With the addition of the terms $a_{1F}$ and $a_{2F}$ we can now consider
the photons of the FBR without assuming them to be experimentally
discriminated and to appreciate the relevance of their contribution to the
RC. In the next section we shall display several numerical evaluations to
illustrate this, both at the level of the DP and at the level of $\alpha_B^T$.

\section{Numerical results}

     In this section we shall perform numerical evaluations of the RC. We have
two purposes in mind. One is to make an internal check of our results and the
other one is to compare with numerical results available in the
literature~\cite{gluck}.

     The internal check consists of performing numerically the photon triple
integrals of Eq.~(\ref{eq:decayrate}) and comparing them with the analytical
result of Eq.~(\ref{eq:dtotal}). This comparison is made over a lattice of
points $(E,E_2)$ of the complete DP. At the same time, the choice of this
lattice is made so as to be able to compare with the numbers of
Ref.~\cite{gluck}. These detailed comparisons will be made specifically for
the decays $\Lambda \rightarrow pe\overline{\nu}$ and $\Sigma^- \rightarrow
ne\overline{\nu}$. It is then necessary that we adopt here the definitions
introduced in Ref.~\cite{gluck} and to take the same values of the
corresponding form factors. Accordingly, we introduce the two-dimensional
function

\begin{equation}
\delta \alpha_B^R(E,E_2) = \alpha_B^R(E,E_2) - \alpha_B^0 (E,E_2) .
\label{eq:alexp}
\end{equation}
$\alpha_B^R(E,E_2)$ and $\alpha_B^0(E,E_2)$ are defined as in
Eq.~(\ref{eq:npnm}), but this time without integrating over $E$ and $E_2$
({\it i.e.} integrating only over $d\Omega_2$). The upper index $R$ has the
same meaning as in Sec.~V.

     When the photon triple integration is to be performed numerically
$\alpha_B^R(E,E_2)$ is explicitly given, according to our discussions of Sec.
II, by

\begin{equation}
\alpha_B^R(E,E_2) = - \frac{A_0^{\prime \prime} + (\alpha/\pi)
\Psi_2}{A_0^\prime + (\alpha/\pi) \Psi_1} , \label{eq:alnum}
\end{equation}
where
\begin{equation}
\Psi_{1} = A_1^\prime (\phi+\theta_1) + A_1^{\prime \prime} \phi^\prime +
\frac{p_2 l}{4\pi} \int_{-1}^1 dx \int_{-1}^{y_0} dy \int_0^{2\pi} d \varphi_k
\left[\left| {\sf M}^\prime \right|^2 + \left| {\sf M}^{\prime \prime}
\right|^2 \right] \end{equation}
and
\begin{equation}
\Psi_{2} = A_2^\prime (\phi+\theta_1) + A_2^{\prime \prime} \phi^\prime +
\frac{p_2l}{4\pi} \int_{-1}^1 dx \int_{-1}^{y_0} dy \int_0^{2\pi} d\varphi_k
\left[ \left| {\sf N}^{\prime \prime \prime} \right|^2 + \left| {\sf N}^{{\rm
IV}} \right|^2 \right] . \end{equation}
The numerical values obtained from our analytical result use

\begin{equation}
\alpha_B^R (E,E_2) = -\frac{A_0^{\prime \prime} + (\alpha/\pi)
\Phi_2}{A_0^\prime + (\alpha/\pi)\Phi_1}, \label{eq:eA}
\end{equation}
where $\Phi_1$ and $\Phi_2$ are given in Eq.~(\ref{eq:dwp}) of Sec.~II. For
the FBR, when the photon triple integration is to be performed numerically, we
introduce the definition
\begin{equation}
\delta \alpha_B^F(E,E_2) = - \frac{\Psi_{2F}}{\Psi_{1F}} , \label{eq:rnum}
\end{equation}
where
\begin{equation}
\Psi_{1F} = A_1^\prime I_{0F}(E,E_2) + \frac{p_2 l}{4\pi} \int_{-1}^1 dx
\int_{-1}^1 dy \int_0^{2\pi} d \varphi_k \left[\left| {\sf M}^\prime
\right|^2 + \left| {\sf M}^{\prime \prime} \right|^2 \right]
\end{equation}
and
\begin{equation}
\Psi_{2F} = A_2^\prime I_{0F}(E,E_2) + \frac{p_2l}{4\pi} \int_{-1}^1 dx
\int_{-1}^1 dy \int_0^{2\pi} d\varphi_k \left[ \left| {\sf M}^{\prime
\prime \prime} \right|^2 + \left| {\sf M}^{{\rm IV}} \right|^2 \right] .
\end{equation}

     The numerical values obtained for the analytical result for the FBR use
the definition
\begin{equation}
\delta \alpha_B^F (E,E_2) = -\frac{\Phi_{2F}}{\Phi_{1F}},
\label{eq:rexp}
\end{equation}
where $\Phi_{2F}$ and $\Phi_{1F}$ are given in Eqs.~(\ref{eq:e66}) and
(\ref{eq:e67}), respectively.

     For $\Lambda \rightarrow pe\overline{\nu}$ and $\Sigma^- \rightarrow
ne\overline{\nu}$ the numerical results are displayed in Tables I and II,
respectively. The lattices in these tables are given in terms of $\delta =
E/E_m$ and $\sigma = E_2/M_1$. In Tables I(a) and II(a) we display the values
obtained with Eqs.~(\ref{eq:alnum}) and (\ref{eq:rnum}). In Tables I(b) and
II(b) we display the values obtained with Eqs.~(\ref{eq:eA}) and
(\ref{eq:rexp}). In Tables I(c) and II(c) the numbers of Ref.~\cite{gluck} are
displayed.

    The internal cross-check in Tables I(a)--II(a) and I(b)--II(b) is very
good. The comparison with Ref.~\cite{gluck} is quite acceptable. Some
minor differences can be observed, but they can reasonably attributed to the
difference in approximations, {\it i.e.}, within our approximations this last
comparison with Ref.~\cite{gluck} is satisfactory.

     As a last step, in Table III we display the values of the totally
integrated spin-asymmetry coefficient $\alpha_B$ for several processes of
interest, namely, $n \rightarrow p e \overline{\nu}$, $\Lambda \rightarrow
pe\overline{\nu}$, $\Sigma^- \rightarrow ne\overline{\nu}$, $\Sigma^-
\rightarrow \Lambda e\overline{\nu}$, $\Sigma^+ \rightarrow \Lambda e^+ \nu$,
$\Xi^- \rightarrow \Lambda e\overline{\nu}$, $\Xi^- \rightarrow \Sigma^0
e\overline{\nu}$, $\Xi^0 \rightarrow \sum^+ e \overline{\nu}$, and
$\Lambda_c^+ \rightarrow \Lambda e^+\nu$. The values of the form factors used
are those given in Ref.~\cite{amv00}. In the second column of this table we
display the uncorrected coefficient $\alpha_B^0$. In the third column we list
the correction to this coefficient defined as \begin{equation} \delta
\alpha_B^R = \alpha_B^R - \alpha_B^0 . \end{equation}
In the next column we list the radiatively corrected $\alpha_B$ for the
complete DP, which is analogously defined as
\begin{equation}
\delta \alpha_B^T = \alpha_B^T - \alpha_B^0.
\end{equation}
In order to compare with our results in the last column we display the values
of $\delta \alpha_B^T$ reported in Ref.~\cite{gluck}.

     From Table III we can appreciate, by comparing the third and the fourth
columns, that the inclusion of the FBR is important. In general it reduces
the total radiative corrections. It may even be that the values in the
third column are one order of magnitude larger than the corresponding ones
of the fourth column. Therefore, there is an important difference
between $\alpha_B^R$ and $\alpha_B^T$. From this Table III we can see that
there is an acceptable agreement between our $\delta \alpha_B^T$ and the
one of Ref.~\cite{gluck} for the two decays reported there. 

\section{Conclusions}

     In this paper we have calculated the RC to the emitted baryon angular
distribution w.r.t. the spin of the decaying baryon of HSD, without the
restriction imposed in Ref.~\cite{rfm97}. This restriction was, that
bremsstrahlung photons be experimentally discriminated either by direct
detection or indirectly by energy-momentum conservation. 

     It proved to be convenient to recast the results of Ref.~\cite{rfm97} in
close parallelism with Ref.~\cite{amv00}, where the emitted charged-lepton
angular distribution w.r.t. the spin of the decaying baryon was studied and
the above restriction was not imposed either. This facilitated our task
greatly in two respects. First, it gave us the differential decay rate with RC
of the TBR of Ref.~\cite{rfm97} in a form that it could be extended to
incorporate the previously discriminated photons of the FBR by simply
replacing the limits of integration over the real photon three-momentum.
Second, it allowed us to express our analytical results using in as-much-as
possible expressions already obtained in Ref.~\cite{amv00}, and thus
considerably reducing the number of new analytical integrals.

     Accordingly, our main result has two very compact forms, given in
Eqs.~(\ref{eq:decayrate}) and (\ref{eq:dtotal}). In the first one the
integrations over the photon three-momentum are explicitly indicated and can
easily be performed numerically. In the second one all such integrations were
performed and a complete analytical result is obtained.

     As an application we computed the RC to $\Lambda \rightarrow pe
\overline{\nu}$ and $\Sigma^- \rightarrow ne\overline{\nu}$ over a detailed
lattice of points covering the TBR and FBR of the DP of these decays. The
results are displayed in Tables I and II, respectively. In these tables we
exhibited an internal cross-check of Eqs.~(\ref{eq:decayrate}) and
(\ref{eq:dtotal}) and a comparison with numerical results published in the
literature \cite{gluck}. The comparisons are satisfactory. In addition, we
calculated the RC to the total asymmetry-coefficient of the emitted baryon
for nine decays, including the charm-baryon decay $\Lambda_c^+ \rightarrow
\Lambda e^+\nu$. The results are displayed in Table III. Here we separated the
contributions of TBR from the RC including also the FBR contributions, and we
also compared with the results for two decays given in Ref.~\cite{gluck}. This
last is also satisfactory within our approximations. The contributions of the
FBR photons to the RC are, generally speaking, quite appreciable and, in some
cases, even reverse the sign of the total RC.

     Our results are useful for a model-independent experimental analysis. They
are reliable up to a precision of around 0.5\% and, thus, are useful for
experiments involving several thousands of events. For high statistics
experiments with several hundreds of thousands of events or for decays
involving charm baryons, such as $\Lambda_c^+ \rightarrow \Lambda e^+\nu$, or
even heavier quarks, our Eqs.~(\ref{eq:decayrate}) and (\ref{eq:dtotal})
provide a good first approximation. To improve the precision of our formulas
it becomes necessary to include terms of order $\alpha q/\pi M_1$. This can be
done still in a model-independent way by extending  the general analysis of
Ref.~\cite{sirlin} for the virtual RC and by use of the Low theorem \cite{low}
for the bremsstrahlung photons.

     We should make a few more remarks. Our results are valid for both
neutral and charged polarized decaying hyperons and whether the
emitted-charged lepton is an electron or a muon. If this lepton is
positively-charged our formulas are also applicable provided the sign of the
$g_i$ form factors and the sign in front of the ${\bf \hat s}_1 \cdot {\bf \hat
p}_2$ correlation are all reversed \cite{torres}. This rule applies equally
well to total asymmetry coefficients $\alpha_B^T$, $\alpha_B^R$, and
$\alpha_B^0$. Finally, let us mention that in a Monte Carlo analysis the
analytical result Eq.~(\ref{eq:dtotal}) represents a considerable advantage,
because the triple photon integration does not have to be repeated every time
the values of $f_1$ and $g_1$, or $E$ and $E_2$ are changed. This leads to a
considerable simplification of the experimental Monte Carlo simulation.

\acknowledgements

     The authors are grateful to Consejo Nacional de Ciencia y
Tecnolog{\'\i}a (M\'exico) for partial support. A.M. \ and J.J.T. \
acknowledge partial support by Comisi\'on de Operaci\'on y Fomento de
Actividades Acad\'emicas (Instituto Polit\'ecnico Nacional).

\begin{table}
\squeezetable
\caption{
Percentage RC $\delta \alpha_B^R(E,E_{2})$ over the TBR and $\delta
\alpha_B^F(E,E_2)$ over the FBR in  $\Lambda \rightarrow pe\overline{\nu }$.
The entries corresponding to the latter are marked with bold-face characters.
The energies $E_2$ and $E$ are replaced by $\sigma = E_2/M_1$ and $\delta =
E/E_m$, respectively. (a) gives the results of the numerical integrations
Eqs.~(\ref{eq:alnum}) and (\ref{eq:rnum}), (b) gives the results of the
analytical formulas Eqs.~(\ref{eq:eA}) and (\ref{eq:rexp}), and (c) gives the
results of Ref.~[3]. In each column we provide the kinematical limits on
$\sigma$ in the TBR in terms of $\sigma^{\rm max}$ and $\sigma^{\rm min}$.}
\begin{tabular}{ddddddddddd}
$\sigma$ & \multicolumn{10}{d}{(a)} \\ \tableline
0.8529 &  0.2 &  0.1 &  0.0 &  0.0 &  0.0 &  0.1 &  0.1 &  0.2 &  0.4 &  1.6 \\
0.8517 &  1.4 &  0.2 &  0.2 &  0.1 &  0.2 &  0.2 &  0.3 &  0.5 &  1.1 &  1.8 \\
0.8504 &\mbox{{\bf -89.8}} &  0.5 &  0.3 &  0.3 &  0.3 &  0.4 &  0.6 &  0.9 &  1.5 &      \\
0.8492 &\mbox{{\bf -85.4}} &  1.0 &  0.6 &  0.5 &  0.5 &  0.6 &  0.8 &  1.2 &  1.6 &      \\
0.8479 &\mbox{{\bf -80.3}} &\mbox{{\bf -93.1}} &  0.9 &  0.7 &  0.7 &  0.8 &  1.1 &  1.4 &  1.4 &
\\
0.8466 &\mbox{{\bf -74.1}} &\mbox{{\bf -81.8}} &  1.3 &  1.0 &  1.0 &  1.1 &  1.3 &  1.5 &  0.5 &
\\
0.8454 &\mbox{{\bf -66.6}} &\mbox{{\bf -73.2}} &  1.9 &  1.4 &  1.3 &  1.4 &  1.5 &  1.5 &      &
\\
0.8441 &\mbox{{\bf -57.3}} &\mbox{{\bf -63.0}} &\mbox{{\bf -74.5}} &  2.0 &  1.7 &  1.7 &  1.7 &
1.0 &  & \\
0.8429 &\mbox{{\bf -45.1}} &\mbox{{\bf -49.7}} &\mbox{{\bf -57.8}} &  2.8 &  2.3 &  2.1 &  1.6 &
&  & \\
0.8416 &\mbox{{\bf -26.5}} &\mbox{{\bf -29.2}} &\mbox{{\bf -33.8}} &\mbox{{\bf -42.5}} &
3.5 &  2.5 & 0.1 & & & \\ \\
& \multicolumn{10}{d}{(b)} \\ \tableline
0.8529 &  0.2 &  0.1 &  0.0 &  0.0 &  0.0 &  0.1 &  0.1 &  0.2 &  0.4 &  1.6 \\
0.8517 &  1.4 &  0.2 &  0.2 &  0.1 &  0.2 &  0.2 &  0.3 &  0.5 &  1.1 &  1.8 \\
0.8504 &\mbox{{\bf -89.8}} &  0.5 &  0.3 &  0.3 &  0.3 &  0.4 &  0.6 &  0.9 &  1.5 &      \\
0.8492 &\mbox{{\bf -85.4}} &  1.0 &  0.6 &  0.5 &  0.5 &  0.6 &  0.8 &  1.2 &  1.6 &      \\
0.8479 &\mbox{{\bf -80.3}} &\mbox{{\bf -93.1}} &  0.9 &  0.7 &  0.7 &  0.8 &  1.1 &  1.4 &  1.4 &
\\
0.8466 &\mbox{{\bf -74.1}} &\mbox{{\bf -81.8}} &  1.3 &  1.0 &  1.0 &  1.1 &  1.3 &  1.5 &  0.5 &
\\
0.8454 &\mbox{{\bf -66.6}} &\mbox{{\bf -73.2}} &  1.9 &  1.4 &  1.3 &  1.4 &  1.5 &  1.5 &      &
\\
0.8441 &\mbox{{\bf -57.3}} &\mbox{{\bf -63.0}} &\mbox{{\bf -74.5}} &  2.0 &  1.7 &  1.7 &  1.7 &
1.0 &  & \\
0.8429 &\mbox{{\bf -45.1}} &\mbox{{\bf -49.7}} &\mbox{{\bf -57.8}} &  2.8 &  2.3 &  2.1 &  1.6 &
&  & \\
0.8416 &\mbox{{\bf -26.5}} &\mbox{{\bf -29.2}} &\mbox{{\bf -33.8}} &\mbox{{\bf -42.5}} &  3.5 &  2.4
& 0.1 & & & \\ \\
\tableline
$\delta$& 0.0500 & 0.1500 & 0.2500 & 0.3500 & 0.4500 & 0.5500 & 0.6500 &
 0.7500 & 0.8500 & 0.9500 \\ \\
        & \multicolumn{10}{d}{(c)} \\ \tableline
0.8530 & 0.2 & 0.1 & 0.1 & 0.1 & 0.1 & 0.1 & 0.1 & 0.2 & 0.3 & 0.9 \\
0.8518 & 1.9 & 0.4 & 0.3 & 0.3 & 0.3 & 0.4 & 0.5 & 0.6 & 1.0 & 1.1 \\
0.8505 &\mbox{{\bf -84.2}} & 0.8 & 0.6 & 0.5 & 0.6 & 0.6 & 0.8 & 1.0 & 1.3
& \\
0.8493 &\mbox{{\bf -77.6}} & 1.4 & 0.9 & 0.8 & 0.8 & 0.9 & 1.1 & 1.3 & 1.4
& \\
0.8480 &\mbox{{\bf -70.5}} & 2.4 & 1.3 & 1.1 & 1.1 & 1.2 & 1.3 & 1.5 & 1.1
& \\
0.8467 &\mbox{{\bf -62.8}} &\mbox{{\bf -72.1}} & 1.7 & 1.4 & 1.4 & 1.5 &
1.6 & 1.5 & 0.3 & \\
0.8455 &\mbox{{\bf -54.4}} &\mbox{{\bf -62.2}} & 2.3 & 1.8 & 1.7 & 1.7 &
1.7 & 1.3 & & \\
0.8442 &\mbox{{\bf -45.2}} &\mbox{{\bf -51.7}}  &\mbox{{\bf -64.6}} & 2.2
& 2.0 & 2.0 & 1.7 & 0.8 & & \\
0.8429 &\mbox{{\bf -34.5}} &\mbox{{\bf -39.5}}  &\mbox{{\bf -47.8}} & 2.8
& 2.5 & 2.2 & 1.5 & & & \\
0.8417 &\mbox{{\bf -20.2}} &\mbox{{\bf -23.1}} &\mbox{{\bf -27.1}}
&\mbox{{\bf -35.2}} & 3.2 & 2.3 & 0.1 & & & \\ \\ \tableline
$\delta$& 0.05 & 0.15 & 0.25 & 0.35 & 0.45 & 0.55 & 0.65 & 0.75 & 0.85 &
0.95 \\ \\
$\sigma^{max}$ & 0.8536 & 0.8536 & 0.8536 & 0.8536 & 0.8536 & 0.8536 & 0.8536 & 0.8536 & 0.8536
& 0.8536 \\
$\sigma^{min}$ & 0.8516 & 0.8479 & 0.8450 & 0.8428 & 0.8414 & 0.8410 & 0.8416 & 0.8433 & 0.8464
& 0.8508 \\ \\
\end{tabular}
\end{table}

\begin{table}
\squeezetable
\caption{
Everything here is as explained in the caption of Table I, except that the
decay studied is now $\Sigma^- \rightarrow n e\overline{\nu}$.
}
\begin{tabular}{ddddddddddd}
$\sigma$ &
\multicolumn{10}{d}{(a)} \\
\tableline
0.8066 &  0.6 &  0.1 &  0.0 &  0.0 &  0.0 &  0.0 &  0.0 &  0.0 &  0.0 &  0.1 \\
0.8043 & 66.0 &  0.3 &  0.1 &  0.1 &  0.0 &  0.0 &  0.0 &  0.0 &  0.1 &  0.3 \\
0.8020 &\mbox{{\bf 54.3}} &  1.2 &  0.3 &  0.2 &  0.1 &  0.1 &  0.0 &  0.1 &  0.1 &
\\
0.7997 &\mbox{{\bf 55.0}} &  5.6 &  0.7 &  0.3 &  0.2 &  0.1 &  0.1 &  0.1 &  0.1 &      \\
0.7973 &\mbox{{\bf 55.0}} &\mbox{{\bf 31.1}} &  1.6 &  0.6 &  0.3 &  0.2 &  0.1 &  0.1 &  0.1 &
\\
0.7950 &\mbox{{\bf 54.5}} &\mbox{{\bf 43.3}} &  4.5 &  1.1 &  0.5 &  0.3 &  0.2 &  0.1 &  0.1 &
\\
0.7927 &\mbox{{\bf 53.1}} &\mbox{{\bf 45.4}} & 20.3 &  2.2 &  0.9 &  0.4 &  0.2 &  0.1 &      &
\\
0.7904 &\mbox{{\bf 50.2}} &\mbox{{\bf 44.5}} &\mbox{{\bf 28.8}} &  5.3 &  1.5
&  0.7 &  0.2 &  0.0 & & \\
0.7881 &\mbox{{\bf 44.1}} &\mbox{{\bf 40.1}} &\mbox{{\bf 31.3}} &\mbox{{\bf
-4.9}} &  3.2 &  1.1 & 0.3 & & & \\
0.7858 &\mbox{{\bf 29.4}} &\mbox{{\bf 27.3}} &\mbox{{\bf 23.2}} &\mbox{{\bf
15.2}} &  9.7 &  2.2 & 0.2 & & & \\ \\
 & \multicolumn{10}{d}{(b)} \\
\tableline
0.8066 &  0.6 &  0.1 &  0.0 &  0.0 &  0.0 &  0.0 &  0.0 &  0.0 &  0.0 &  0.1 \\
0.8043 & 66.0 &  0.3 &  0.1 &  0.1 &  0.0 &  0.0 &  0.0 &  0.0 &  0.1 &  0.3 \\
0.8020 &\mbox{{\bf 54.3}} &  1.2 &  0.3 &  0.2 &  0.1 &  0.1 &  0.1 &  0.1 &  0.1 &      \\
0.7997 &\mbox{{\bf 55.0}} &  5.6 &  0.7 &  0.3 &  0.2 &  0.1 &  0.1 &  0.1 &  0.1 &      \\
0.7973 &\mbox{{\bf 55.0}} &\mbox{{\bf 31.1}} &  1.6 &  0.6 &  0.3 &  0.2 &  0.1 &  0.1 &  0.1 &
\\
0.7950 &\mbox{{\bf 54.5}} &\mbox{{\bf 43.3}} &  4.5 &  1.1 &  0.5 &  0.3 &  0.2 &  0.1 &  0.1 &
\\
0.7927 &\mbox{{\bf 53.1}} &\mbox{{\bf 45.4}} & 20.3 &  2.2 &  0.9 &  0.4 &  0.2 &  0.1 &      &
\\
0.7904 &\mbox{{\bf 50.2}} &\mbox{{\bf 44.5}} &\mbox{{\bf 28.7}} &  5.3 &  1.5 &  0.7 &  0.3 &  0.0 &
& \\
0.7881 &\mbox{{\bf 44.1}} &\mbox{{\bf 40.1}} &\mbox{{\bf 31.2}} &\mbox{{\bf -5.2}} &  3.2 &  1.1 &
0.3 & & & \\
0.7858 &\mbox{{\bf 29.4}} &\mbox{{\bf 27.3}} &\mbox{{\bf 23.0}} &\mbox{{\bf 14.3}} &  9.8 &  2.2 &
0.2 & & & \\ \\ \tableline
$\delta$& 0.0500 & 0.1500 & 0.2500 & 0.3500 & 0.4500 & 0.5500 & 0.6500 &
 0.7500 & 0.8500 & 0.9500 \\ \\
        & \multicolumn{10}{d}{(c)} \\ \tableline
0.8067 & 0.6 & 0.1 & 0.0 & 0.0 & 0.0 & 0.0 & 0.0 & 0.0 & 0.0 & 0.1 \\
0.8044 & 50.7 & 0.3 & 0.1 & 0.0 & 0.0 & 0.0 & 0.0 & 0.0 & 0.1 & 0.2 \\
0.8020 &\mbox{{\bf 60.7}} & 1.2 & 0.2 & 0.1 & 0.1 & 0.0 & 0.0 & 0.1 & 0.1
& \\
0.7997 &\mbox{{\bf 62.4}} & 5.4 & 0.6 & 0.2 & 0.1 & 0.1 & 0.1 & 0.1 & 0.2
& \\
0.7974 &\mbox{{\bf 63.7}} &\mbox{{\bf 47.3}} & 1.4 & 0.4 & 0.2 & 0.1 &
0.1 & 0.1 & 0.2 & \\
0.7951 &\mbox{{\bf 64.5}} &\mbox{{\bf 56.7}} & 4.0 & 0.8 & 0.3 & 0.2 & 0.1
& 0.1 & 0.1 & \\
0.7928 &\mbox{{\bf 64.5}} &\mbox{{\bf 58.7}} & 18.4 & 1.7 & 0.6 & 0.3 &
0.2 & 0.1 & & \\
0.7904 &\mbox{{\bf 62.8}} &\mbox{{\bf 58.1}}  &\mbox{{\bf 45.1}} & 4.4 &
1.0 & 0.4 & 0.2 & 0.1 & & \\
0.7881 &\mbox{{\bf 57.1}} &\mbox{{\bf 53.6}}  &\mbox{{\bf 45.7}} & 11.1 &
2.4 & 0.7 & 0.2 & & & \\
0.7858 &\mbox{{\bf 39.8}} &\mbox{{\bf 37.8}} &\mbox{{\bf 33.6}}
&\mbox{{\bf 25.1}} & 8.2 & 1.4 & 0.1 & & & \\ \\ \tableline
$\delta$& 0.0500 & 0.1500 & 0.25 & 0.35 & 0.45 & 0.55 & 0.65 &
 0.75 & 0.85 & 0.95 \\ \\
$\sigma^{max}$ & 0.8078 & 0.8078 & 0.8078 & 0.8078 & 0.8078 & 0.8078 & 0.8078 & 0.8078 & 0.8078
& 0.8078 \\
$\sigma^{min}$ & 0.8043 & 0.7978 & 0.7925 & 0.7884 & 0.7857 & 0.7846 & 0.7854 & 0.7884 & 0.7938
& 0.8023 \\ \\
\end{tabular}
\end{table}

\begin{table}[h]
\caption{Percentage RC to the total spin-asymmetry coefficient of the emitted
baryon for nine HSD. RC of the TBR have been separated, in the third column,
from the total RC including the FBR photons displayed in the fourth column. In
these calculations the analytical result Eq.~(\ref{eq:dtotal}) was employed.
The last column reproduces the numerical results of Ref.~[3].
} 
\begin{tabular}{ldddd}
Decay & $\alpha_B^0$ & $\delta \alpha_B^R = \alpha_B^R - \alpha_B^0$ & $ \delta
\alpha_B^T = \alpha_B^T - \alpha_B^0$ & $\delta \alpha_B^T$ Ref.~[3] \\ 
\tableline
$n\rightarrow p$ & $-$47.92 & $-$0.28 & $-$0.28 & \\
$\Lambda \rightarrow p$ &-58.60 & $-$0.20 & $-$0.26 & $-$0.1 \\
$\Sigma^-\rightarrow n$ & 66.73 & 0.12 & $-$0.03 & $-$0.0 \\
$\Sigma^-\rightarrow \Lambda$ & 7.24 & 0.12 & $-$0.12 & \\
$\Sigma^+\rightarrow \Lambda$ & 6.59 & $-$0.05 & 0.10 & \\
$\Xi^-\rightarrow \Lambda$ & $-$54.72 & 0.04 & $-$0.01 & \\
$\Xi^-\rightarrow \Sigma^0$ & $-$45.87 & $-$0.01 & $-$0.09 & \\
$\Xi^0\rightarrow \Sigma^+$ & $-$46.15 & $-$0.16 & $-$0.23 & \\
$\Lambda_c^+\rightarrow \Lambda$ &$-$31.06 & $-$0.93 & 0.12 & \\
\end{tabular}
\end{table}

\end{document}